\documentclass{new_tlp}
\usepackage{mathptmx}
\usepackage{graphicx}

\newcommand\bcmdtab{\noindent\bgroup\tabcolsep=0pt%
  \begin{tabular}{@{}p{10pc}@{}p{20pc}@{}}}
\newcommand\ecmdtab{\end{tabular}\egroup}

\usepackage{xspace}
\usepackage[ruled,linesnumbered,vlined]{algorithm2e}\SetProcNameSty{textsc}
\usepackage{subfigure}
\usepackage{epsfig}
\usepackage{epstopdf}
\usepackage{comment}
\usepackage{amssymb}
\usepackage{amsmath}
\usepackage[T1]{fontenc}
\usepackage{url}
\usepackage{graphicx,amssymb}
\usepackage[textsize=small]{todonotes}
\usepackage{booktabs}
\usepackage{multirow}
\usepackage{tikz}
\usepackage{pgfplots}\usetikzlibrary{plotmarks}
\usepackage{textcomp}
\usepackage{pdfpages}

\usepackage{listings}
\makeatletter
\lst@Key{countblanklines}{true}[t]{\lstKV@SetIf{#1}\lst@ifcountblanklines}
\lst@AddToHook{OnEmptyLine}{%
    \lst@ifnumberblanklines\else%
    \lst@ifcountblanklines\else%
    \advance\c@lstnumber-\@ne\relax%
    \fi%
    \fi}
\makeatother
\lstdefinelanguage{asp}{
    breakatwhitespace=true,
    captionpos=b,
    numbers=left,
    numbersep=5pt,
    numberblanklines=false,
    countblanklines=false,
    commentstyle=\color{gray},
    frame=bt, framexbottommargin=5pt, framextopmargin=5pt,
    aboveskip=5pt, belowskip=5pt,
    abovecaptionskip=10pt
}
\lstnewenvironment{asp}{
	\lstset{
		language=asp,
		showstringspaces=false,
        keepspaces=true,
		formfeed=\newpage,
		tabsize=4,
		numbers=none,
		breaklines=true,
		literate={~} {$\sim$}{1},
    frame=none,
	}
}{}

\usepackage[strings]{underscore}

\newcommand{\rev}[1]{\color{#1}\xspace}

\newtheorem{example}{Example}[section]

  \title[ValAsp: a tool for data validation in Answer Set Programming]
        {ValAsp: a tool for data validation in Answer Set Programming}

  \author[M. Alviano and C. Dodaro and A. Zamayla]
         {
             MARIO ALVIANO and CARMINE DODARO AND ARNEL ZAMAYLA\\
             Department of Mathematics and Computer Science, University of Calabria, Italy\\
	Via P. Bucci, cubo 30B, 87036, Rende (CS), Italy\\
	\email{\{alviano,dodaro,zamayla\}@mat.unical.it}\\
	}

\jdate{March 2003}
\pubyear{2003}
\pagerange{\pageref{firstpage}--\pageref{lastpage}}
\doi{S1471068401001193}

\begin{document}

\label{firstpage}

\maketitle

  \begin{abstract}
The development of complex software requires tools promoting fail-fast approaches, so that bugs and unexpected behavior can be quickly identified and fixed. Tools for data validation may save the day of computer programmers. In fact, processing invalid data is a waste of resources at best, and a drama at worst if the problem remains unnoticed and wrong results are used for business. Answer Set Programming (ASP) is not an exception, but the quest for better and better performance resulted in systems that essentially do not validate data. Even under the simplistic assumption that input/output data are eventually validated by external tools, invalid data may appear in other portions of the program, and go undetected until some other module of the designed software suddenly breaks. This paper formalizes the problem of data validation for ASP programs, introduces a language to specify data validation, and presents \textsc{valasp}, a tool to inject data validation in ordinary programs. The proposed approach promotes fail-fast techniques at coding time without imposing any lag on the deployed system if data are pretended to be valid.
Validation can be specified in terms of statements using YAML, ASP and Python. Additionally, the proposed approach opens the possibility to use ASP for validating data of imperative programming languages.
Under consideration for acceptance in TPLP.
   \end{abstract}



\section{Introduction}

A popular Latin saying starts with \emph{errare humanum est} (translated, to err is human), and clarifies how making mistakes is part of human nature.
Computer programmers, being humans, are inclined and not indifferent to errors \cite{DBLP:conf/vl/KoM03,DBLP:journals/vlc/KoM05}.
Whether a typo in notation, a misspelled word, or a wrong or fragile representation of data, errors in source code files may result in substantial delays in developing an application.
Even worse, errors may stay unknown for a long time, until something happens that stimulates the error to cause a crash of the system or some other unwanted and unexpected behavior.
In the worst scenario, unknown errors may lead to wrong results that are used to take some business decision, which in turn may ruin a company.
(Refer to the paper by \citeN{DBLP:journals/tse/NatellaWCS20} for examples of typical errors in software systems.)

Fail-fast systems are designed to break as soon as an unexpected condition is detected (refer to the paper by \citeN{DBLP:conf/oopsla/PadhyeS19} for an example of fail-fast type checking in Java).
As often it is the case, the idea is not limited to computer programming, and definitely not originated in software design.
For example, many electrical devices have fail-fast mechanisms to provide overcurrent protection --- it is better to melt an inexpensive fuse, than to burn a control board.
Even if technically an electrical device operating with a fuse replaced by a wire works as expected in most cases, no professional electrician would suggest to tamper with the system in this way.
In fact, in case the replaced fuses melt again and again, it is usually evidence that the system has some malfunction that must be detected and fixed.

Computer programming should follow a similar fail-fast approach.
Errors must be detected as soon as possible, and reported to the programmer in a non-skippable way, so that the malfunction can be quickly detected and fixed.
Data validation is the process of ensuring that data conform to some specification, so that any process in which they are involved can safely work under all of the assumptions guaranteed by the specification.
In particular, immutable objects are usually expected to be validated on creation, so that their consistency can be safely assumed anywhere in the system --- this way, in fact, an invalid immutable object simply cannot exist because its construction would fail.
Guaranteeing validity of mutable objects is usually more difficult and tedious, and almost impossible if there is no invariant on the validity of immutable objects \rev{black}that are part of the mutable objects\rev{black}.
(Refer to the papers by \citeN{SecureByDesign} and \citeN{DDDD} for details on how to tackle the complexity of a business domain in terms of domain primitives and entities.)

Answer Set Programming (ASP; \citeNP{DBLP:journals/ngc/GelfondL91,niem-99,mare-trus-99}) should not be an exception when it comes to errors.
However, the language offers very little in terms of protection mechanisms.
No static or dynamic type checking are available, and programmers can rely on a very limited set of primitive types, namely integers, strings and alphanumeric constants, with no idiomatic way to enforce that the values of an argument must be of one of these types only --- refer to the paper by \citeN{DBLP:journals/tplp/CalimeriFGIKKLM20} for details of the ASP-Core-2 format.
More structured data types are usually represented by uninterpreted function symbols \cite{DBLP:conf/iclp/LierlerL09,DBLP:conf/ijcai/EiterS09,DBLP:journals/tplp/BaseliceB10,DBLP:journals/aicom/CalimeriCIL11,DBLP:journals/tplp/AlvianoFL10}, but again there is no idiomatic way to really validate such structures.
Similarly, there is no idiomatic way to specify that input relations satisfy some properties (often expressed in comments, to document the usage of ASP encodings).
Even \emph{integrity constraints} may be insufficient to achieve a reliable data validation, as they are cheerfully satisfied if at least one of their literals is false;
in fact, integrity constraints are very convenient for discarding unwanted solutions, but not very effective in guaranteeing data integrity --- invalid data in this case may lead to discarding some wanted solution among thousands or more equally acceptable solutions, a very hard-to-spot unexpected outcome.

The lack of data validation in ASP is likely due to the quest for better and better performance.
After a significant effort to optimize the search algorithms that are one of the main reasons of the success of ASP systems and solvers like \textsc{clingo} \cite{DBLP:journals/ki/GebserKKLOORSSW18}, \textsc{dlv} \cite{DBLP:conf/lpnmr/LeoneAACCCFFGLC19} and \textsc{wasp} \cite{DBLP:conf/cilc/DodaroAFLRS11}, to sacrifice a few machine instructions just to validate data that are almost always valid sounds like a blasphemy.
Someone may argue that ASP is not intended to be a general purpose programming language, and therefore input and output are eventually validated by external tools.
However, this is a very simplistic assumption, and invalid data may appear in other portions of the program, causing the loss of otherwise acceptable solutions.
Everyone is free to follow their own path, but at some point in their life, perhaps after spending hours looking for a typo in the name of a function, any programmer will regret not having had an idiomatic way to specify the format of their data.
Quoting the Latins, \emph{errare humanum est, perseverare autem diabolicum} --- to err is human, but to persist (in error) is diabolical.

This paper aims at rescuing ASP programmers from some problems due to data validation by proposing a framework called \textsc{valasp}, written in Python and available at \url{https://github.com/alviano/valasp}.
Specifically, a first contribution of this paper is the formalization of the problem of data validation for ASP programs by combining ASP rules with Python exceptions (Section~\ref{sec:data-validation}).
A second contribution of this paper is a language based on the YAML serialization format to specify data validation for ASP programs (Section~\ref{sec:yaml-layer}), and its compilation into Python code that can be processed by the ASP system \textsc{clingo} (Section~\ref{sec:python-layer}).
The proposed approach is to specify the format of some data of interest, leaving open the possibility to work with unspecified data types (Section~\ref{sec:framework}).
Moreover, the specification can be separated from the ASP program, and the fail-fast approach is achieved by injecting constraints that are guaranteed to be immediately satisfied when grounded, unless data are found to be invalid and some exception is raised.
Such a behavior is obtained thanks to \emph{interpreted functions}, briefly recalled in Section~\ref{sec:background}, where their use for data validation is also hinted.
Finally, a few use cases are discussed in Section~\ref{sec:usecases}, among them the possibility to take advantage of ASP declarativity for validating complex Python data structures, and related work from the literature is discussed in Section~\ref{sec:rw} --- in particular, differences with sort typed systems like IDP \cite{DBLP:books/mc/18/Cat0BJD18} and SPARC \cite{DBLP:journals/tplp/MarcopoulosZ19}.
\footnote{An extended abstract of this work was presented at the International Conference on Logic Programming (ICLP) 2020 \cite{Alviano2020}.}

This paper extends a previous work presented at PADL 2021 \cite{DBLP:conf/padl/AlvianoDZ21} by providing more detailed examples of the implemented approach, by presenting additional use cases of the proposed validation framework, by showing how \textsc{valasp} can be used to validate Python data structures, and by reporting an extended empirical assessment.

\section{Background}\label{sec:background}

ASP programs are usually evaluated by a two-steps procedure:
first, object variables are eliminated by means of \emph{intelligent grounding} techniques, and after that stable models of the resulting propositional program are searched by means of sophisticated non-chronological backtracking algorithms.
Details of this procedure, as well as on the syntax and semantics of ASP, are out of the scope of this work.
Therefore, this section only recalls the minimal background required to introduce the concepts presented in the next sections.

ASP is not particularly rich in terms of primitive types, and essentially allows for using integers and (double-quoted) strings.
(We will use the syntax of \textsc{clingo} \cite{DBLP:journals/ki/GebserKKLOORSSW18}.)
More complex types, as for example dates or decimal numbers, can be represented by means of (non-interpreted) functions, or by the so called \emph{@-terms};
in the latter case, the @-term must be associated with a function (written in an imperative programming language like Python) mapping different objects to different symbols in a repeatable way --- for example, by populating a table of symbols or by using a natural enumeration.

\begin{example}[Primitive types and @-terms]\label{ex:date}
Dates can be represented by strings, functions (tuples as a special case) or @-terms, among other possibilities.
Hence, \lstinline|"1983/09/12"|, \lstinline|date(1983,9,12)| and \lstinline|@date(1983,9,12)| can all represent the date 12 Sep.\ 1983, where the @-term is associated with the following Python code:
\begin{asp}
def date(year, month, day):
    res = datetime.datetime(year.number, month.number, day.number)
    return int(res.timestamp())
\end{asp}
Each representation comes with pros and cons, discussed later in Section~\ref{sec:framework}.
\hfill$\blacksquare$
\end{example}

Intelligent grounding may process rules in several orders, and literals within a rule can also be processed according to different orderings.
A safe assumption made here is that all object variables of an @-term must be already bound to some ground term before the grounder can call the associated (Python) function.

\begin{example}[@-term invocation]\label{ex:triple}
Consider the following program:
\begin{asp}
birthday(sofia, date(2019,6,25)).
birthday(bigel, date(1982,123)).  % Oops! I missed a comma, but where?!?
:- birthday(Person,Date), @is_triple_of_integer(Date) != 1.
\end{asp}
The Python function associated with the @-term is called two times, with arguments \lstinline|date(2019,6,25)| and \lstinline|date(1982,123)|, so that some invariant can be enforced on the second argument of every instance of \lstinline|birthday/2|.
\hfill$\blacksquare$
\end{example}

Data validation is the process of ensuring that data conform to some specification, so that any process in which they are involved can safely work under all of the assumptions guaranteed by the specification.
Data can be found invalid because of an expected error-prone source (for example, user input from a terminal), or due to an unexpected misuse of some functionality of a system (this is usually the case with bugs).
While in the first case it is natural to ask again for the data, in the second case failing fast may be the only reasonable choice, so that the problem can be noticed, traced, and eventually fixed.
The fail-fast approach is particularly helpful at coding time, to avoid bug hunting at a later time, but it may also pay off after deployment if properly coupled with a recovery mechanism (for example, restart the process).

\begin{example}[Data validation]\label{ex:triple-fun}
The @-term from Example~\ref{ex:triple} can be associated with the following Python code:
\begin{asp}
def is_triple_of_integer(value):
    if value.type != Function: raise ValueError('wrong type')
    if value.name != 'date': raise ValueError('wrong name')
    if len(value.arguments) != 3: raise ValueError('not a triple')
    if any(arg for arg in value.arguments if arg.type != Number):
        raise ValueError('arguments must be integers')
    return 1
\end{asp}
Indeed, the presence of \lstinline|birthday(bigel, date(1982,123))| will be noticed because of abrupt termination of the grounding procedure.
Adopting a fail-fast approach is the correct choice in this case, and any attempt of sanification is just a dangerous speculation on the invalid data --- should it be \lstinline|date(1982,1,23)|, or \lstinline|date(1982,12,3)|?
\hfill$\blacksquare$
\end{example}

\section{A data validation framework for ASP}\label{sec:framework}

Data validation can be used in ASP programs thanks to @-terms.
However, the resulting code is likely to be less readable due to aspects that are not really the focus of the problem aimed to be addressed.
We will illustrate our proposal to accomplish data validation without cluttering an ASP encoding in this section.
First, the problem of data validation for ASP programs is formalized in Section~\ref{sec:data-validation}, and a few minimal examples are provided.
After that, a language based on the YAML serialization format is introduced in Section~\ref{sec:yaml-layer} to specify data validation for ASP programs.
Finally, Section~\ref{sec:python-layer} illustrates how the YAML format is compiled into Python code that can be processed by the ASP system \textsc{clingo}.

\subsection{Data validation for ASP programs}\label{sec:data-validation}

Let us fix a set $\mathcal{R}$ of \emph{predicate and function names} (or \emph{symbols}), a set $\mathcal{U}$ of \emph{field names}, and a set $\mathcal{T} = \{$\lstinline|Integer|, \lstinline|String|, \lstinline|Alpha|, \lstinline|Any|$\}$ of \emph{primitive types}.
Each type is associated with a set of \emph{facets}, or restrictions.
The facets of \lstinline|Integer| are 
\lstinline|enum| to specify a list of acceptable values,
\lstinline|min| (by default $-2^{31}$) and
\lstinline|max| (by default $2^{31}-1$) to specify (inclusive) bounds, and finally
\lstinline|count|, \lstinline|sum+| and \lstinline|sum-| to specify bounds on the number of values, the sum of positive values and negative values.
The facets of \lstinline|String| and \lstinline|Alpha| are 
\lstinline|enum| and \lstinline|count| as before,
\lstinline|min| and \lstinline|max| to bound the length, and
\lstinline|pattern| to specify a regular expression.
Other types have only the facet \lstinline|count|.

A \emph{user-defined symbol} $s$ is any name in $\mathcal{R}$.
A \emph{field declaration} is a tuple of the form $(f,t,F)$, where $f$ is a field name in $\mathcal{U}$, $t$ is a type in $\mathcal{T}$ or a type defined by the user (i.e., a user-defined symbol), and $F$ is a set of facets for $t$.
A \emph{field comparison} is an expression of the form $f \odot f'$, where $\odot$ is a comparison operator among \lstinline|==|,
\lstinline|!=|,
\lstinline|<|,
\lstinline|<=|,
\lstinline|>=|, and
\lstinline|>|.

A \emph{user definition} is a tuple of the form $(s,D,H,b,c,a)$, where $s$ is a user-defined symbol, $D$ is a set of field declarations, $H$ is a set of field comparisons (also called \emph{having properties}), and $b,c,a$ are code blocks to be executed respectively before grounding, after the creation of an instance of the user-defined symbol, and after grounding.
A \emph{data validation specification} is a tuple of the form $(P,A,U)$, where $P$ is a code block, $A$ is an ASP program, and $U$ is a set of user definitions.

\begin{example}[Validation of dates]\label{ex:birthday}
Let \lstinline|date| be a ternary predicate representing a valid date, and \lstinline|bday| be a binary predicate whose arguments represent a person and a date.
A user definition $u_{date}$ of \lstinline|date| could be \lstinline|(date, {(year, Integer, $\emptyset$), (month, Integer, $\emptyset$), (day, Integer, $\emptyset$)}, $\emptyset$, $\emptyset$, $c$, $\emptyset$)|, where $c$ is the following (Python) code block:
\begin{asp}
datetime.datetime(self.year, self.month, self.day).
\end{asp}
A user definition $u_{bday}$ of \lstinline|bday| could be \lstinline|(bday, {(name, Alpha, $\emptyset$), (date, date, $\emptyset$)}, $\emptyset$, $\emptyset$, $\emptyset$, $\emptyset$)|.
A data validation specification could be the triple $(P,$ $\emptyset,$ $\{u_{date}, u_{bday}\})$ where $P$ is the (Python) code block
\begin{asp}
import datetime
\end{asp}
\vspace{-1.75em}
\hfill$\blacksquare$
\end{example}

\begin{example}[Ordering of elements]\label{ex:order}
Let \lstinline|ordered_triple| be a ternary predicate representing a triple of integers in descendent order.
A user definition could be \lstinline|(ordered_triple,| \lstinline|{(first, Integer, $\emptyset$),| \lstinline|(second, Integer, $\emptyset$),| \linebreak \lstinline|(third, Integer, $\emptyset$)},| \lstinline|{first < second,| \lstinline|second < third}, $\emptyset$, $\emptyset$, $\emptyset$)|.
\hfill$\blacksquare$
\end{example}

\begin{example}[Overflow on integers]\label{ex:income}
Let \lstinline|income| be a binary predicate representing incomes of companies, which are summed up in an ASP program.
A user definition of \lstinline|income| could be \lstinline|$u_\mathit{income} :=$ (income, {(company, String, $\emptyset$), (amount, Integer, {min: 0, sum+: 2147483647})}, $\emptyset$, $\emptyset$, $\emptyset$, $\emptyset$)|, specifying that valid \linebreak amounts are nonnegative and their sum must not overflow.
A data validation specification could be $(\emptyset,\emptyset,\{u_\mathit{income}\})$.
\hfill$\blacksquare$
\end{example}

\begin{example}[Validation of complex aggregates]\label{ex:overflow}
Consider the constraint
\begin{asp}
:- bound(MAX), #sum{B-B',R : init_budget(R,B), budget_spent(R,B')} > MAX.
\end{asp}
It bounds the total amount of residual budget, for example for researchers involved in a project.
The above constraint can be part of a broader ASP program where \lstinline|budget_spent/2| depends on some guess on resources and services to purchase.
The aggregate above may overflow, and we are interested in detecting such cases and stopping the computation on such unreliable data.
To this aim, we can introduce auxiliary predicates in the ASP program $A$ of a data validation specification $(\emptyset,A,U)$:
\begin{asp}
residual_budget(B-B',R) :- init_budget(R,B), budget_spent(R,B').
\end{asp}
Hence, we can provide a user definition \lstinline|(residual_budget, {(value, Integer, {min: 0, sum+: 2**31-1}), (id_res, Integer, {min: 0})}, $\emptyset$, $\emptyset$, $\emptyset$)| in $U$.
\hfill$\blacksquare$
\end{example}

\paragraph{Specification application.}
Given an ASP program $\Pi$, and a data validation specification $(P,A,U)$, the application of $(P,A,U)$ to $\Pi$ amounts to the following computational steps:
\begin{enumerate}
\item
The code block $P$ is executed.

\item
For all $(s,D,H,b,c,a) \in U$, the code block $b$ is executed.

\item
The ASP program $\Pi \cup A$ is grounded.

\item
For all produced instances of a predicate $s$ such that $(s,D,H,b,c,a) \in U$, all types and facets in $D$ and all field comparisons in $H$ are checked, and the code block $c$ is executed.
If a check fails, an exception is raised.

\item
For all $(s,D,H,b,c,a) \in U$, the code block $a$ is executed.
\end{enumerate}

\begin{example}
The application of the data validation specification from Example~\ref{ex:birthday} to an ASP program whose intelligent grounding produces \lstinline|bday(bigel, date(1982,123))| raises an exception due to the wrong type of the second argument, that is, function \lstinline|date| is expected to have arity 3, but only 2 arguments are found.

The application of the data validation specification from Example~\ref{ex:income} to an ASP program comprising facts
\lstinline|income("Acme ASP",1500000000)| and \lstinline|income("Yoyodyne YAML",1500000000)| raises an exception due to the facet \lstinline|sum+: 2147483647| of \lstinline|amount|. 
This way, an overflow is prevented, for example in
\begin{asp}
total(T) :- T = #sum{A,C : income(C,A)}.
\end{asp}
which would otherwise produce \lstinline|total(-1294967296)| in \textsc{clingo} and \textsc{dlv}.
\hfill$\blacksquare$
\end{example}

\subsection{A YAML language for data validation}\label{sec:yaml-layer}

YAML is a human friendly data serialization standard, whose syntax is well-suited for materializing the notion of data validation specification provided in Section~\ref{sec:data-validation}.
The YAML files processed by our framework are essentially dictionaries associating keys to other dictionaries, values, and lists.
The key \lstinline|valasp| is reserved, and cannot be used as a symbol or field name.
 \rev{black}Finally, code blocks are written in Python.\rev{black}

More in details, a data validation specification $(P,A,U)$ is represented by a YAML file comprising the following lines:
\begin{asp}
valasp:
    python: |+
        <Python code block $P$>
    asp: |+
        <ASP program $A$>
\end{asp}
and a block of lines for each user definition $(s,D,H,b,c,a)$:
\begin{asp}
$s$:
    <field declarations $D$>    
    valasp:
        having:
            - <field comparison $h_1$>
            - ...
            - <field comparison $h_n$>
        before_grounding: |+
            <Python code block $b$>
        after_init: |+
            <Python code block $c$>
        after_grounding: |+
            <Python code block $a$>
\end{asp}
Above, $h_1,\ldots,h_n$ are the field comparisons in $H$ (for some $n \geq 0$), and a field declaration $(f,t,F)$ is represented by
\begin{asp}
    $f$:
        type: $t$
        <facets $F$>
\end{asp}
where facets are written as key-value pairs.

\begin{example}\label{ex:yaml}
Below is a YAML file to validate predicate \lstinline|bday| of Example~\ref{ex:birthday}.
\begin{asp}
valasp:
    python: |+
        import datetime

date:
    year: Integer
    month: Integer
    day: Integer    

    valasp:
        after_init: |+
            datetime.datetime(self.year, self.month, self.day)

bday:
    name: Alpha
    date: date
\end{asp}
The following, instead, is a YAML file to validate predicate \lstinline|ordered_triple| of Example~\ref{ex:order}:
\begin{asp}
ordered_triple:
	first: Integer
	second: Integer
	third: Integer
	
	valasp:
		having:
			- first < second
			- second < third
\end{asp}
Note that YAML lists can be written as multiple lines starting with a dash, or in square brackets.	
Regarding predicate \lstinline|income| of Example~\ref{ex:income}, its YAML file is the following:
\begin{asp}
income:
    company: String
    amount:
        type: Integer
        min: 0
        sum+: Integer
\end{asp}
Here, \lstinline|sum+: Integer| is syntactic sugar for specifying that the sum of positive values must fit into a 32-bits integer --- nicer than writing \lstinline|max: 2147483647| or \lstinline|max: 2**31-1| inside \lstinline|sum+|. 
\hfill$\blacksquare$
\end{example}


%
%
%
%

\subsection{The Python compilation}\label{sec:python-layer}

The specification for data validation expressed in YAML is compiled into Python code that can be processed by the ASP system \textsc{clingo}.
The compilation injects data validation in the grounding process by introducing \emph{constraint validators} of two kinds, namely \emph{forward} and \emph{implicit}, depending on the arity of the validated predicates and on how terms are passed to @-terms:
for unary predicates, their unique terms are forwarded directly to the functions handling @-terms;
for other predicates, instead, terms are grouped by functions with the same name of the validated predicate.
Hence, for a predicate \lstinline|pred| of arity $1$, the (forward) constraint validator has the following form:
\begin{asp}
:- pred(X1), @valasp_validate_pred(X1) != 1.
\end{asp}
Similarly, for a predicate \lstinline|pred| of arity $n \geq 2$, the (implicit) constraint validator has the following form:
\begin{asp}
:- pred(X1,...,X$n$), @valasp_validate_pred(pred(X1,...,X$n$)) != 1.
\end{asp}
In both cases, @-terms are associated with the Python function
\begin{asp}
def valasp_validate_pred(value):
    Pred(value)
    return 1
\end{asp}
where \lstinline|Pred| is a class whose name is obtained by capitalizing the first lowercase letter of \lstinline|pred|, and whose constructor raises an exception if the provided data are invalid.
In fact, class \lstinline|Pred| is also an outcome of the compilation process, and materializes all validity conditions specified in the data validation specification in input.

In a nutshell, given a data validation specification $(P,A,U)$ (represented in YAML  \rev{black}and whose code blocks are written in Python\rev{black}), and an ASP program $\Pi$, the compilation produces a Python script with the following content:
\begin{enumerate}
\item
The Python program $P$.

\item
A Python class $S$ for every $(s,D,H,b,c,a) \in U$ materializing all validity conditions:
field declarations in $D$ map to Python class annotations (and added as instance attributes on instance creation);
field comparisons in $H$ and the Python code block $c$ are added to the \lstinline|__post_init__| method (and executed after any instance creation);
the Python code blocks $b$ and $a$ are respectively added to the class methods \lstinline|before_grounding| and \lstinline|after_grounding|.

\item
Calls to any \lstinline|before_grounding| method introduced in the previous steps.

\item
Calls to \textsc{clingo}'s API to ground the ASP program $\Pi \cup A \cup C$, where $C$ is the set of constraint validators associated with $U$.

\item
Calls to any \lstinline|after_grounding| method introduced in the previous steps.
\end{enumerate}

\begin{example}[Continuing Example~\ref{ex:yaml}]\label{ex:python}
The YAML file to validate predicate \lstinline|bday| of Example~\ref{ex:birthday} is mapped to the following Python code:
\begin{asp}
import datetime

context = Context(wrap=[])

@context.valasp
class Date:
   	year: Integer
   	month: Integer
   	day: Integer
       
   	def __post_init__(self):
   		datetime.datetime(self.year, self.month, self.day)
    		
@context.valasp
class Bday:
   	name: Alpha
   	date: Date
\end{asp}
The two Python classes, \lstinline|Date| and \lstinline|Bday|, are decorated with the decorator \lstinline|@context.valasp|, which is in charge for interpreting the annotations used to declare fields:
the constructor of (the decorated) \lstinline|Date| class checks the presence of three integer arguments, namely \lstinline|year|, \lstinline|month| and \lstinline|day|, and calls the \lstinline|__post_init__| method to ensure that they form a valid date;
the constructor of (the decorated) \lstinline|Bday| class checks that the first argument is alphanumeric and the second argument is a valid instance of \lstinline|Date|.

The YAML file to validate predicate \lstinline|ordered_triple| of Example~\ref{ex:order} is mapped to the following Python code:
\begin{asp}
@context.valasp
class Ordered_triple:
   	first: Integer
   	second: Integer
   	third: Integer
       
   	def __post_init__(self):
   		if not self.first < self.second:
           raise ValueError("Expected first < second")
   		if not self.second < self.third:
           raise ValueError("Expected second < third")
\end{asp}
Note that the two \lstinline|having| constraints are mapped to two conditional statements, and comprehensive messages are provided in case of violation.

Finally, the YAML file to validate predicate \lstinline|income| of Example~\ref{ex:income} is mapped to the following Python code:
\begin{asp}
@context.valasp
class Income:
   	company: String
   	amount: Integer

   	def __post_init__(self):
   		if self.amount < 0:
           raise ValueError(f"Should be >= 0, but received {self.amount}")
   		if self.amount > 0:
   			self.__class__.sum_positive_of_amount += self.amount

   	@classmethod
   	def before_grounding_init_positive_sum_amount(cls):
       cls.sum_positive_of_amount = 0

   	@classmethod
   	def after_grounding_check_positive_sum_amount(cls):
   		if cls.sum_positive_of_amount > 2147483647:
           raise ValueError('sum of amount in income may exceed 2147483647')
\end{asp}
Above we can observe that the \lstinline|min| constraint is enforced in the first line of \lstinline|__post_init__|, while the \lstinline|sum+| constraint requires to initialize the class variable \lstinline|sum_positive_of_amount| (in method \lstinline|before_grounding_init_positive_sum_amount|), to update its value for each positive \lstinline|amount| (in \lstinline|__post_init__|), and a final check (in method \lstinline|after_grounding_check_positive_sum_amount|). \rev{black}Note that integers in Python do not have fixed byte length. Therefore, there is no overflow in \lstinline|sum_positive_of_amount|.
\rev{black}

The decorated classes are then used for validation by means of the following code:
\begin{asp}
control = clingo.Control()
control.load("-")
try:
    context.valasp_run(control,
        on_validation_done=lambda: print("ALL VALID!\n=========="), 
        on_model=lambda m: print(f"Answer: {m}\n=========="), 
    )
except RuntimeError as e:
    raise ValueError(context.valasp_extract_error_message(e)) from None
\end{asp}
Code similar to the above snippet is also produced by the translation of YAML files.
As an alternative, validation can be specified directly in terms of the above Python classes, and the programmer can customize the invocation of \textsc{valasp}.
\hfill$\blacksquare$
\end{example}

\section{Use cases and assessment}\label{sec:usecases}

This section reports a few use cases on two encodings from ASP competitions \cite{DBLP:journals/tplp/GebserMR20}.
Each use case focuses on the validation of parts of an encoding, showing how the proposed framework can identify invalid data. Note that tuning of the encoding is out of the scope of this paper.
Moreover, the overhead introduced by data validation is empirically assessed.
Finally, an application of \textsc{valasp} for validating complex Python data structures is shown.

\subsection{Video streaming --- 7th ASP competition} 
Video streaming amounts to selecting an optimal set of video representations, in terms of resolution and bit-rate, to satisfy user requirements.
User requirements and solution are respectively encoded by \lstinline|user(USERID, VIDEOTYPE, RESOLUTION, BANDWIDTH, MAXSAT, MAXBITRATE)| and \lstinline|assign(USER_ID, VIDEO_TYPE, RESOLUTION, BITRATE, SAT)|.
The overall satisfaction of users is maximized by the following weak constraint:
\begin{asp}	
:~ assign(USER_ID,_,_,BITRATE,SAT_VALUE), user(USER_ID,_,_,_,BEST_SAT,_).
   [BEST_SAT-SAT_VALUE@1, USER_ID, assign]
\end{asp}
According to the official description, available online at \url{http://aspcomp2015.dibris.unige.it/Video_Streaming.pdf}, instances of \lstinline|user/6| can be validated with the following YAML specification:
\begin{asp}
user:
    userid:
    	type: Integer
    	min: 0
    videotype:
    	type: String
    	enum: [Documentary, Video, Cartoon, Sport]    
    resolution:
    	type: Integer
    	enum: [224, 360, 720, 1080]
    bandwidth:
    	type: Integer
    	min: 0
    maxsat:
        type: Integer
    	min: 0
    maxbitrate:
        type: Integer
    	min: 150
    	max: 8650        
    valasp:
    	after_init: |+
    	   if self.maxbitrate % 50 != 0: raise ValueError("unexpected bitrate")
\end{asp}
According to the above specification, the arguments \lstinline|userid|, \lstinline|bandwidth| and \lstinline|maxsat| are non-negative integers;
\lstinline|videotype| is a string among \lstinline|Documentary|, \lstinline|Video|, \lstinline|Cartoon|, and  \lstinline|Sport|; 
argument \lstinline|resolution| is an integer among 224, 360, 720, and 1080;
and \lstinline|maxbitrate| is an integer between 150 and 8650, and it is divisible by 50.

The official encoding and instances do not have errors, as expected.
However, the encoding is quite fragile and relies on several assumptions on the input data and on ASP internals --- ASP systems use 32-bits integers for everything but the cost of a solution.
To show how dangerous such assumptions are, consider a decision problem where a partial solution and a target satisfaction are given.
Accordingly, the weak constraint is replaced by the following constraint:
\begin{asp}
:- target(T), #sum{BEST_SAT-SAT_VALUE, USER_ID : 
    assign(USER_ID,_,_,BITRATE,SAT_VALUE), user(USER_ID,_,_,_,BEST_SAT,_)} > T.
\end{asp}
In this case, the execution of \textsc{clingo} on the instances of the competition may lead to the error message
\lstinline|"Value too large to be stored in data type: Integer overflow!"|, produced while simplifying the sum.
However, whether the message is shown or not depends on the partial solution provided in input.
In fact, if the overflow is only due to the \lstinline|assign/5| instances in input, the subsequent simplification step cannot notice the problem and a wrong answer is produced.
 \rev{black}For example, if  \lstinline|BEST_SAT=1500000000| and the input contains two \lstinline|assign/5| instances with \lstinline|SAT_VALUE=1| and \lstinline|SAT_VALUE=2|, then the grounder of \textsc{clingo} will simplify the aggregate by communicating to the solver that the minimum value is \lstinline|(1500000000-1+1500000000-2)$\ $ mod $2^{31}$=852516349|, which is interpreted as \lstinline|$2^{31}$-852516349=-1294967299|; at this point, if the other values associated to the undefined instances of \lstinline|assign/5| are not sufficient to overflow the sum, then the previous overflow is unnoticed.\rev{black}
The following YAML specification can help to detect these overflows:
\begin{asp}
target:
    value:
      type: Integer
      min: 0
sum_element:
    value:
        type: Integer
        min: 0
        sum+: Integer
   userid: Integer
valasp:
    asp: |+
        sum_element(BEST_SAT-SAT_VALUE,UID) :- 
            assign(UID,_,_,BITRATE,SAT_VALUE), user(UID,_,_,_,BEST_SAT,_).	
\end{asp}

\subsection{Solitaire --- 4th ASP Competition} 
Solitaire represents a single-player game played on a 7x7 board where the 2x2 corners are omitted.
We focus on the following rules defining the board:
\begin{asp}
range(1).
range(X+1) :- range(X), X < 7.
location(1,X) :- range(X), 3 <= X, X <= 5.
location(2,X) :- range(X), 3 <= X, X <= 5.
location(Y,X) :- range(Y), 3 <= Y, Y <= 5, range(X).
location(6,X) :- range(X), 3 <= X, X <= 5.
location(7,X) :- range(X), 3 <= X, X <= 5.
\end{asp}
Those rules are interesting since an error in this point might be propagated all over the encoding.
The YAML specification of \lstinline|range| and \lstinline|location| is the following:
\begin{asp}
range:
    value:
        type: Integer
        enum: [1, 2, 3, 4, 5, 6, 7]
location:
    x: range
    y: range    
    valasp:
        after_grounding: |+
            pos = [1,2,6,7]
            if self.x.value in pos and self.y.value in pos:
                raise ValueError("Invalid position")
\end{asp}

 \rev{black}In particular, note that this example shows the usage of the \lstinline|after_grounding| statement to check the valid positions after that all grounding instances of \lstinline|location| have been created.\rev{black}

\subsection{Qualitative spatial reasoning --- 4th ASP Competition}
Qualitative spatial reasoning consists of deciding whether a set of spatial and temporal constraints is consistent with respect to a composition table.
Membership in qualitative relations is encoded by 169 rules, similar to the following:
\begin{asp}
	label(X,Z,rp) :- label(X,Y,rp), label(Y,Z,rp).
	label(X,Z,req) | label(X,Z,rp) | label(X,Z,rpi) | label(X,Z,rd) | label(X,Z,rdi)
	| label(X,Z,rs) | label(X,Z,rsi) | label(X,Z,rf) | label(X,Z,rfi) 
	| label(X,Z,rm) | label(X,Z,rmi) | label(X,Z,ro) | label(X,Z,roi)
	:- label(X,Y,rp), label(Y,Z,rpi).
\end{asp}
The third argument of \lstinline|label/3| is a qualitative relation.
The following YAML specification can be used to validate such rules:
\begin{asp}
rel:
  value:
    type: Alpha
    enum: [req, rp, rpi, rd, rdi, ro, roi, rm, rmi, rs, rsi, rf, rfi]
node:
  value:
    type: Integer
    min: 0
    max: 49
label:
  x: node
  y: node
  l: rel
  valasp:
    having: [x < y]
\end{asp}

 \rev{black}The above example shows the usage of the \lstinline|having| statement to compare the values of the two fields of the predicate \lstinline|label|.\rev{black}

\subsection{Knight's Tour --- 3rd ASP Competition} 
The knight's tour problem aims at finding a sequence of moves of a knight on a chessboard of size $N$ such that the knight visits every square exactly once and comes back to the origin. We focus on a short excerpt of the encoding\footnote{The full version can be found at \url{http://www.mat.unical.it/aspcomp2011/files/KnightTour/knight_tour-full_package.zip}}:
\begin{asp}
size(8). givenmove(7,5,8,7). givenmove(1,7,3,6). 

number(X) :- size(X).
number(X) :- number(Y), X=Y-1, X>0.
even :- size(N), number(X), N = X+X.
:- not even.
:- size(N), N < 6.
\end{asp}
Those rules are particularly interesting from the point of view of the validation.
First of all, the first line contains a test case, probably this was a test used by a programmer that was not commented out before publication. Note that atoms of the form givenmove/4 are part of the input, therefore the ones added in the encoding might lead to incorrect results. 
Moreover, remaining rules are used to check that the size of the chessboard must be an even number greater than 6, which we argue should not be part of the encoding. 

The YAML specification of \lstinline|size| is the following:
\begin{asp}
size:
  value:
    type: Integer
	  min: 6
	  max: 100
	  count: 1
	
  valasp:
    after_init: |+
	  if self.value %2 != 0:
	    raise ValueError('Size must be an even number')
	  self.__class__.value = self.value
\end{asp}

In this case the validation will fail since there are two different atoms of the form size/1. In addition, the following YAML specification of \lstinline|move| and  \lstinline|givenmove| can be used to validate knight's moves:
\begin{asp}
valasp:
  asp: |+
    __in_range(X1, givenmove(X1,Y1,X2,Y2)) :- givenmove(X1,Y1,X2,Y2).
    __in_range(Y1, givenmove(X1,Y1,X2,Y2)) :- givenmove(X1,Y1,X2,Y2).
    __in_range(X2, givenmove(X1,Y1,X2,Y2)) :- givenmove(X1,Y1,X2,Y2).
    __in_range(Y2, givenmove(X1,Y1,X2,Y2)) :- givenmove(X1,Y1,X2,Y2).

    __in_range(X1, move(X1,Y1,X2,Y2)) :- move(X1,Y1,X2,Y2).
    __in_range(Y1, move(X1,Y1,X2,Y2)) :- move(X1,Y1,X2,Y2).
    __in_range(X2, move(X1,Y1,X2,Y2)) :- move(X1,Y1,X2,Y2).
    __in_range(Y2, move(X1,Y1,X2,Y2)) :- move(X1,Y1,X2,Y2).

__in_range:
  x:
    type: Integer
    min: 1
    source: Any

valasp:
  before_grounding: |+
    cls.post_check = []

  after_init: |+
    self.__class__.post_check.append(self)

  after_grounding: |+
    for el in cls.post_check:
      if el.x > Size.value:
        raise ValueError(f'Value out of bound in {el.source}: {el.x}')
\end{asp}
(Note that the above example uses f-strings; \url{https://www.python.org/dev/peps/pep-0498/}.)

\subsection{Empirical assessment}

The overhead introduced by \textsc{valasp} to validate instances of the discussed problems was measured by running \textsc{clingo} with and without validation.
The experiment was run on a 2.4 GHz Quad-Core Intel Core i5 with 16 GB of memory.
\textsc{valasp} was executed with the command-line option \lstinline|--valid-only|, and \textsc{clingo} was executed with its Python interface;
in both cases we disabled the computation of stable models since \textsc{valasp} has no impact on the solving procedure.
We remark here that the running time of \textsc{valasp} includes grounding time.
For each benchmark, we considered all available instances. \rev{black}Results are reported in Table~\ref{tab:experiments}.\rev{black}

\begin{table}[h]
	 \rev{black}
	\begin{tabular}{lrrr}
		Benchmark & \# & \textsc{clingo} & \textsc{valasp}\\
		\cmidrule{1-4}
		Video streaming & 43 & 0.06 & 0.18 \\
		Solitaire & 27 & 0.07 & 0.13 \\
		Qualitative spatial reasoning & 159 & 3.23 & 3.45\\
		Knight tour & 10 & 0.27 &  0.50 \\
	\end{tabular}
	\caption{Average running time (in seconds) of \textsc{clingo} and \textsc{valasp} on tested benchmarks.\label{tab:experiments}}
	\rev{black}
\end{table}
Concerning video streaming, the average running time of \textsc{clingo} is 0.06 seconds, and the average running time of \textsc{valasp} is 0.18 seconds.
As for Solitaire, the average running time of \textsc{clingo} and \textsc{valasp} is respectively 0.07 and 0.13 seconds.
Concerning qualitative spatial reasoning, the average running time of \textsc{clingo} is 3.23 seconds, and the average running time of \textsc{valasp} is 3.45 seconds.
Finally, on knight's tour,  the average running time of \textsc{clingo} and \textsc{valasp} is respectively 0.27 and 0.50 seconds.


We can conclude that no significative overhead is eventually introduced by \textsc{valasp} on these testcases.

\subsection{Application: \textsc{valasp} to validate Python data}

\textsc{valasp} is not only a framework for the validation of ASP data, but also brings the declarative power of ASP to validate complex Python data.
For example, consider a Python function $F$ receiving in input a partially ordered set, that is, a binary relation being reflexive, symmetric, and transitive.
The binary relation is stored in a Python data structure, for example a list of pairs or a sparse matrix.
The Python function $F$ works on the provided data under the assumption that it represents a partially ordered set.
If input data is properly validated, the Python function should verify that the binary relation is actually reflexive, symmetric, and transitive.
Usually, such a validation is achieved by implementing Python functions, with imperative and error-prone code.
\textsc{valasp} provides an alternative:
the relation in input $R$ can be mapped to ASP facts of the form \lstinline|r(a,b)|, for all $(a,b) \in R$, for example with the help of a library like \textsc{clorm} (\url{https://github.com/potassco/clorm}), and the following data validation specification can be used:
\begin{asp}
valasp:
    asp: |+
        element(X) :- r(X,Y). 
        element(Y) :- r(X,Y). 
        lost("reflexivity", X) :- element(X), not r(X,X).
        lost("symmetry", (X,Y)) :- r(X,Y), not r(Y,X).
        lost("transitivity", (X,Y,Z)) :- r(X,Y), r(Y,Z), not r(X,Z).

lost:
    property: String
    reason: Any
    valasp:
        after_init: |+
            raise ValueError(f"Lost {self.property} on {self.reason}")
\end{asp}
If relation $R$ is not a partially ordered set, then it misses at least one property among reflexivity, symmetry, and transitivity.
Such a knowledge is encoded in the ASP program above, which eventually produces an instance of \lstinline|lost/2|.
According to the above data validation specification, \textsc{valasp} will then execute the Python code block given in the \lstinline|after_init|, thus raising an exception to inhibit the execution of function $F$ on invalid data.

As another example of this kind, consider a Python function receiving in input an undirected graph and working under the assumption that the graph is connected.
In order to validate such a precondition, the input graph can be mapped to the ASP predicates \lstinline|vertex/1| and \lstinline|edge/2|,  and the following data validation specification can be used:
\begin{asp}
valasp:
    asp: |+
        connected(FIRST) :- FIRST = #min{X : node(X)}.
        connected(Y) :- connected(X), edge(X,Y).
        unconnected(X) :- node(X), not connected(X).

unconnected:
    node: Any
    valasp:
        after_init: |+
            raise ValueError(f"Unconnected node {self.node}")
\end{asp}
If the input graph is not connected, an exception is raised, pointing to the unconnected node.

\paragraph{Empirical assessment.}
In order to evaluate the performance of \textsc{valasp} to validate Python data we considered an implementation of the Prim's algorithm for computing a minimum spanning tree~\cite{6773228}.
Indeed, a minimum spanning tree can be computed only for connected graphs, therefore we can use the validation presented above to check whether the input graph of the Python function is connected.
We executed the experiment on instances of the problem Graceful Graphs submitted to the 7th ASP Competition, since they all contain connected graphs. Moreover, we associated each edge of the graph with a positive weight.
For each instance, we executed the Python function with and without validation. In the first case, the average running time was 0.06 seconds, whereas in the second case it was 0.05 seconds.

\section{Related work}\label{sec:rw}

The use of types in programming languages eases the representation of complex knowledge, favors the early detection of errors and provides an implicit documentation of source codes \cite{DBLP:books/daglib/0005958}.
For example, by stating that the arguments of predicate \lstinline|bday| are of types \lstinline|person_name| and \lstinline|date|, there is no need to document the way these elements are represented, and any attempt to instantiate this predicate with different types is blocked.
ASP-Core-2 \cite{DBLP:journals/tplp/CalimeriFGIKKLM20}, on the other hand, is untyped:
there is no way to state that arguments of a predicate must be of a specific type, the language offers a very limited set of primitive types, and there is no idiomatic way to declare user-defined types.
This work targets ASP-Core-2,  \rev{black}the standardized language implemented by \rev{black} \textsc{clingo} \cite{DBLP:journals/ki/GebserKKLOORSSW18} and \textsc{dlv2} \cite{DBLP:journals/ki/AdrianACCDFFLMP18}, aiming at providing the missing idioms to specify types and to validate data.

Types are not new in logic-based languages, and in particular order-sorted logic has been formalized as first-order logic with sorted terms, where sorts are ordered to build a hierarchy \cite{DBLP:journals/ai/Kaneiwa04}.
\textsc{idp3} \cite{DBLP:books/mc/18/Cat0BJD18} and \textsc{sparc} \cite{DBLP:journals/tplp/MarcopoulosZ19} are two systems with languages close to ASP-Core-2 and supporting sorted terms.
There are many differences between these systems and the framework proposed in this work.
First of all, \textsc{valasp} is designed to be smoothly integrated with ASP-Core-2 projects:
the programmer is free to choose what to validate and what to leave unchecked, and the original encoding can still be used as it is in case validation is not required in the deployed software.
Sorted terms are also used to bound object variables in rules, while this is not possible with \textsc{valasp} because it only deals with the aspect of data validation.

The framework uses @-terms to perform data validation by means of Python functions that are called during the grounding process.
In the literature, @-terms and non-Herbrand functions \cite{DBLP:journals/tplp/Balduccini13} were used to enrich ASP with functionality that are otherwise not viable (if not in the Turing tarpit).
External atoms in \textsc{hex} \cite{DBLP:journals/ki/EiterGIKRSW18} extend the notion of externally interpreted function to externally interpreted relations, and can be also used to achieve some form of data validation \cite{DBLP:conf/padl/Redl17}.

Intuitively, the constraints
\begin{asp}
:- pred(X1), @valasp_validate_pred(X1) != 1.
:- pred(X1,...,X$n$), @valasp_validate_pred(pred(X1,...,X$n$)) != 1.
\end{asp}
can be replaced by the following HEX rule:
\begin{asp}
:- &valasp_validate_pred[pred]().
\end{asp}
The implementation of the external atom \lstinline|valasp_validate_pred| is similar to the implementation of the @-term \lstinline|valasp_validate_pred|:
it must call the constructor of the (decorated) class \lstinline|Pred| produced by \textsc{valasp} and return the empty relation;
if the construction of \lstinline|Pred| fails, \textsc{valasp} raises an error and blocks the grounding of the program, and otherwise the empty relation returned by the external atom is such that the above constraint is satisfied.
Hence, external atoms can be used as an alternative to @-terms for implementing the validation constraints defined in Section~\ref{sec:python-layer}.

Finally, there are works in the literature that introduce data validation in Prolog systems \cite{kiel1991tool} and that implement data validation for Constraint Logic Programming by means of Prolog systems \cite{DBLP:conf/icalp/HermenegildoPBL02,DBLP:conf/discipl/PueblaBH00a}.
The goal of those works is clearly related to this paper, but they differ on the way data validation is specified, on the target language and on the underlying implementation.
Similarly, debugging techniques for ASP \cite{DBLP:journals/tplp/FandinnoS19,DBLP:conf/aaai/GebserPST08,DBLP:journals/tplp/OetschPT10,DBLP:journals/tplp/DodaroGRRS19} share the goal to identify errors, but with a different approach.
\textsc{valasp} aims at blocking data validation errors in a very early stage, at coding time and by implementing fail-fast techniques to point to the source of the problem.
Debugging techniques instead are useful to localize the origin of unintended behavior, and usually require interaction with the programmer.
If \textsc{valasp} is properly used, a debugger is still a useful software in the tool belt of an ASP programmer, but on the other hand it is likely that the number of debugging sessions will be reduced.
 \rev{black}Moreover, \textsc{valasp} is non-intrusive, since it does not require any change to the tested ASP program, differently from other recently-proposed techniques \cite{DBLP:conf/jelia/AmendolaBR21,DBLP:journals/tplp/Lifschitz17}.\rev{black}

\section{Conclusion}\label{sec:conclusion}

ASP programmers do mistakes, there is no shame in this.
\textsc{valasp} aims at early detection of data validity errors, and promotes a fail-fast approach so that the origin of the problem can be quickly identified and fixed.
The proposed approach follows the separation of concerns design principle:
validation rules are specified in YAML with Python and ASP snippets, and are separated from the business logic represented in ASP encodings.
Such a design is useful to smoothly introduce data validation in ASP, as validation rules can be specified externally without the need to deeply change the way programs are written.
If after deployment data can be safely assumed valid, \textsc{valasp} can be easily discharged because the original encoding stays unchanged.
Moreover, \textsc{valasp} opens the possibility to take advantage of ASP declarativity for validating complex Python data structures, bringing the expression of data validation specifications at a higher level of abstraction.

 \rev{black}
Finally, albeit \textsc{valasp} has a tight integration with the state-of-the-art solver \textsc{clingo}, it can already be used in combination with other ASP solvers based on the ASP-Core-2 standardized language, e.g., \textsc{dlv}. In particular, a user can:
\begin{itemize}
\item use \textsc{valasp} for validation, and \textsc{dlv} for execution during the development phase; or 
\item use \textsc{valasp} for validation and grounding, and \textsc{dlv} for stable model search (e.g. by using the option \texttt{----mode=wasp}) during the deployment phase; or
\item use \textsc{valasp} only during the development phase for early detection of bugs, and then use \textsc{dlv} without validation during the deployment phase.
\end{itemize}
Moreover, it is important to observe that \textsc{valasp} is mainly based on @-terms, a feature that is implemented in \textsc{clingo} but still unavailable in \textsc{dlv}. This is currently the most difficult technical aspect to overcome in order to integrate \textsc{valasp} in  \textsc{dlv}.
\rev{black}

\textbf{Competing interests:} The authors declare none.

\bibliographystyle{acmtrans}
\bibliography{bibliography}

\label{lastpage}
\end{document}